\documentclass[proceedings]{JHEP3}

\PrHEP{PrHEP hep2001}                   
\conference{International Europhysics Conference on HEP}  

\usepackage{epsfig,cite}
\def\beq{\begin{equation}}
\def\eeq{\end{equation}}
\def\bea{\begin{eqnarray}}
\def\eea{\end{eqnarray}}

\def\xbj{x_{\mbox{\scriptsize Bj}}}
\def\sss{\scriptscriptstyle}

\def\rho{\varrho}
\def\pom{I\!\!P}
\title{Non-Perturbative High-Energy QCD}

\author{Arthur Hebecker\\
        Theory Division, CERN, CH-1211 Geneva 23, Switzerland\\      
        E-mail: \email{arthur.hebecker@cern.ch}}

\abstract{It is the aim of this talk to review our understanding of the 
high-energy limit of QCD, focussing, in particular, on recent theoretical 
developments. After a brief introduction, I will recall why the true 
high-energy limit of QCD scattering processes is genuinly non-perturbative 
and why it has so far not been possible to apply lattice methods to this 
type of physics. Given the experimental fact of slowly rising hadronic 
cross sections, we are thus faced with a fundamental problem comparable to 
that of confinement but without the promise of the lattice. During the 
last years, the experimental side of this field has largely been driven by 
the HERA accelerator, which has, naturally, also influenced recent 
theoretical work in high-energy QCD. I will therefore devote the second 
part of the talk to small-$x$ deep inelastic scattering, in particular the 
physics of diffraction, and attempt to describe its impact on the wider 
field of non-perturbative high-energy QCD.}

\begin{document}

\section{Introduction}

\EPSFIGURE[ht]{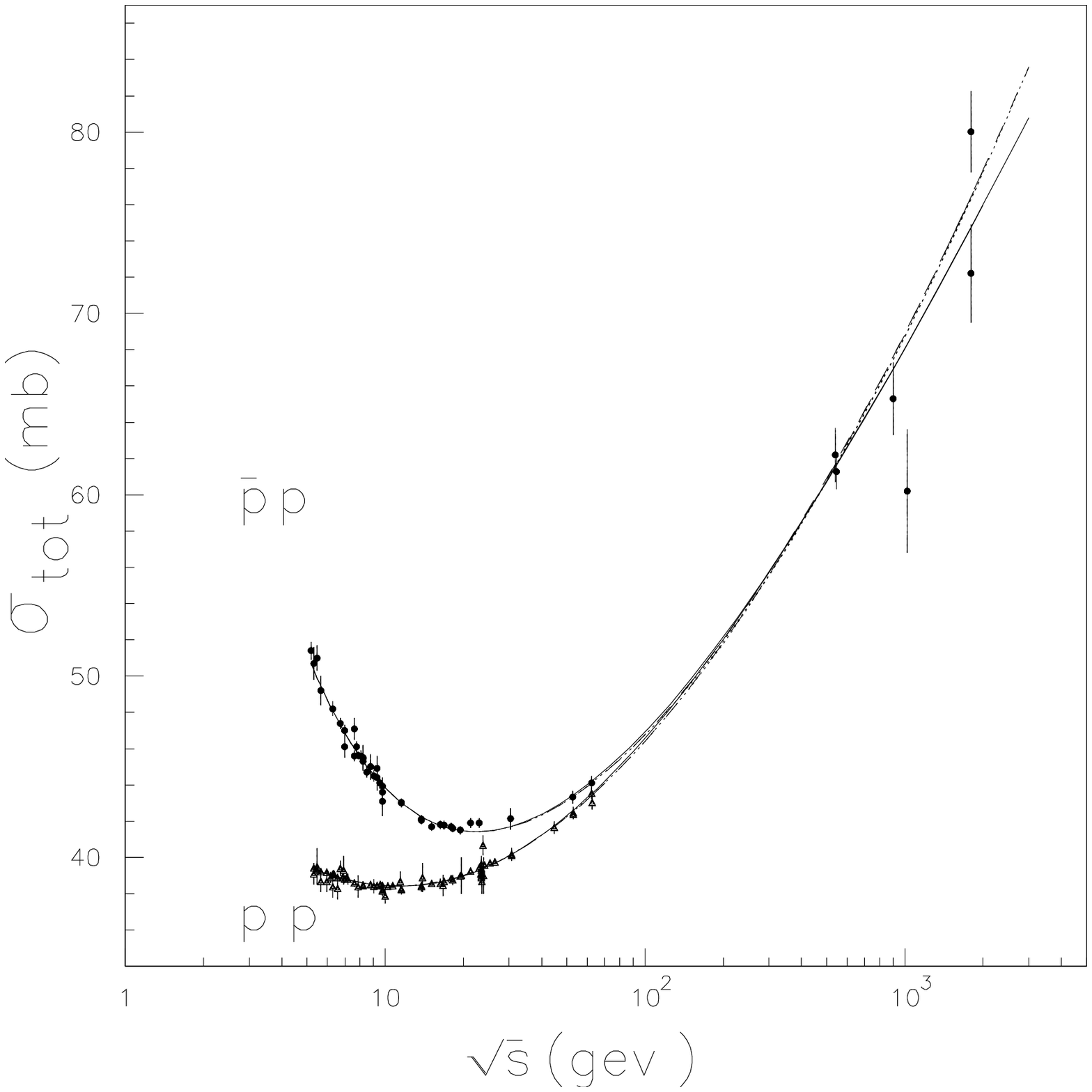,width=0.59\linewidth}
{Total cross sections (from Ref.~\cite{dgm}). For a more recent 
analysis see, e.g.,~\cite{cud}.\label{dgm}}

On the basis of an overwhelming amount of experimental data and its 
quantitative description by various theoretical methods, we can confidently
say that quantum chromodynamics (QCD) is the correct theory of strong 
interactions. Structurally, this theory is extremely simple. It is defined 
by the gauge group SU(3) and the presence of a certain number of quark 
fields $\psi$ in the fundamental representation,
\beq
{\cal L}=-\frac{1}{2}\mbox{tr}F_{\mu\nu}F^{\mu\nu}+i\bar{\psi}(iD\!\!\!\!/
-m)\psi\,. 
\eeq
The coupling constant $\alpha_s=g^2/(4\pi)$ enters the above lagrangian 
via $D_\mu=\partial_\mu+igA_\mu$ and $F_{\mu\nu}=(1/ig)[D_\mu,D_\nu]$. 
The renormalization group teaches us that $\alpha_s$ is small in 
short-distance (i.e., high-virtuality) processes and grows to large values 
in calculations relevant to long-distance phenomena (see, e.g.,~\cite{esw}). 
Thus, while the first type of processes can be controlled in perturbation 
theory, genuinely non-perturbative methods are needed to describe the physics 
of hadrons and low-energy interactions. QCD is therefore arguably the 
best-defined and at the same time most interesting example of a quantum
field theory that we know. The latter statement refers to the physics of
strong interactions and the effect of confinement, which govern the 
low-energy domain. For a large and constantly growing number of low-energy 
observables, numerical simulations on the lattice, i.e., in a discretised 
euclidean version of the theory, provide a first-principles quantitative 
method of calculation (see, e.g.,~\cite{wit}).

However, there exists a qualitatively distinct class of observables, namely 
scattering processes with large center-of-mass energies but without large 
virtualities in intermediate states, where both perturbation theory and 
conventional euclidean lattice methods fail. These processes are the 
focus of the present review. Maybe the most prominent representative of 
these processes are total hadronic cross sections in the limit $s\to\infty$. 
The most interesting qualitative property of these cross sections is their 
slow rise with $s$, which is illustrated in Fig.~\ref{dgm}. The two cross 
sections ($pp$ and $p\bar{p}$) can be parameterized by $s^\delta$ (here 
$\delta=0.07$), $\ln s$ and $\ln^2 s$ at high energy. These 
parameterizations are at present indistinguishable on the basis of the data.
It is surprising and highly unsatisfactory that we have no genuine 
understanding of the striking and universal phenomenon of asymptotically 
rising cross sections on the basis of the QCD lagrangian.

\section{Theoretical ideas concerning the asymptotic rise of cross sections}
\subsection{Geometrical picture}
Let us start with the simple observation that the high-energy total cross 
section of two hadrons (i.e., two complex extended objects) should, naively, 
be constant. They will simply always interact if they overlap in impact 
parameter space. 

It is amusing to note that a very simple quantum-mechanical extension of 
this naive picture was given by Heisenberg as early as 1952~\cite{hei}. He 
assumed that the target hadron is surrounded by a field with energy 
density $\sim e^{-m_\pi r}$ (motivated by the Yukawa potential). Furthermore 
he conjectured that for an inelastic process to occur, the projectile has 
to pass so close to the target that there is {\it locally} enough energy in 
the collision of projectile and target field to create a pion pair. As the
projectile energy grows, this effective maximal impact parameter grows as 
well, and one finds $\sigma\sim(1/m_\pi^2)\ln^2(s/m_\pi^2)$ for the total 
cross section. However, I am not aware of any quantum field theoretic 
version of the above argument. 

Later it was shown on the basis of very general principles, such as 
unitarity and analyticity, that cross sections can not grow faster than 
$\ln^2(s/s_0)$ in the limit $s\to\infty$~\cite{fro} (Froissart bound). 
However, in contrast to Heisenberg's simple argument, the derivation of 
this rigorous results does not provide a physical mechanism realising this 
growth. Thus, there is at present no clear field theoretic understanding 
of the transverse `expansion' of hadrons at high energies.

\subsection{Regge theory}
Let us recall the basic underlying concepts of Regge theory, which,
although not linked directly to the QCD lagrangian, provides a well-defined 
framework for the discussion of high-energy cross sections (see, 
e.g.,~\cite{col,fr}). 

\EPSFIGURE[ht]{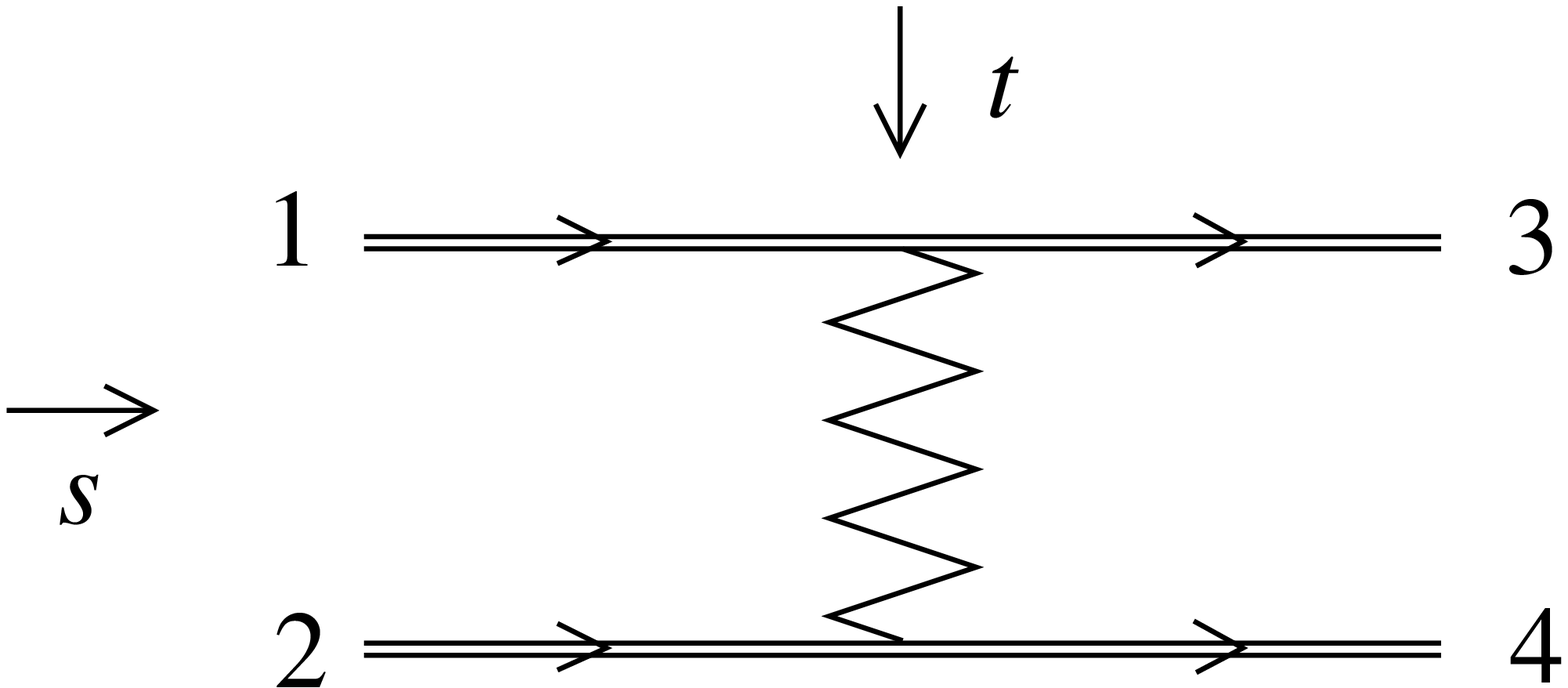,width=0.4\linewidth}
{Scattering process $12\to 34$ via reggeon exchange.\label{reg}}

Using analyticity and crossing symmetry, the amplitude $T_{12\to 34}(s,t)$, 
depicted in Fig.~\ref{reg}, can be related to the amplitude $T_{1\bar{3} 
\to\bar{2}4}(s',t')$, where $s'=t$, $t'=s$, and barred numbers denote 
antiparticles. The partial wave expansion for this crossed amplitude reads 
\beq
T_{1\bar{3}\to\bar{2}4}(s',t')=\sum_{l=0}^{\infty}(2l+1)a_l(s')P_l
(\cos\theta)\,,
\eeq
where $\theta$ is the centre-of-mass frame scattering angle, which is a 
function of $s',t'$ and the particle masses, and $P_l$ are Legendre 
polynomials. Let the two functions $a_\eta(l,t)$ with $\eta=+1$ and 
$\eta=-1$ be the analytic continuations to complex $l$ of the two sequences 
$\{a_l(t),\,\,l=0,2,4,...\}$ and $\{a_l(t),\,\,l=1,3,5,...\}$. In the 
simplest non-trivial case, the only singularity of $a_\eta(l,t)$ is a single 
$t$-dependent pole at $l=\alpha(t)$. It can then be shown that, in the limit 
$s\to\infty$,
\beq
T_{12\to 34}(s,t)=\beta_{13}(t)\beta_{24}(t)\,\zeta_\eta(\alpha(t))\,
\left(\frac{s}{s_0}\right)^{\alpha(t)}\,,\label{rf}
\eeq
where $s_0$ is an arbitrary scale factor, $\beta_{13}$ and $\beta_{24}$ are 
two unknown functions of $t$, and 
\beq
\zeta_\eta(\alpha(t))=\frac{1+\eta e^{-i\pi\alpha(t)}}{\sin \pi\alpha(t)}
\label{sifa}
\eeq
is the signature factor, depending on the signature $\eta$ of the relevant 
Regge trajectory $\alpha(t)$. If $a_\eta(l,t)$ has a more complicated 
analytic structure, the rightmost singularity in the $l$ plane dominates 
the behaviour at large $s$. 

Within the present context, the essential predictions of the asymptotic 
expression Eq.~(\ref{rf}) are the power-like energy dependence 
$s^{\alpha(t)}$ and the factorization of the two vertex factors 
$\beta_{13}(t)$ and $\beta_{24}(t)$. This last feature, 
which underlies the graphic representation of reggeon exchange in 
Fig.~\ref{reg}, is relevant if the same Regge trajectory governs 
different scattering processes. Note also that, for positive $t=s'$ and 
integer $l$, $\alpha(t)$ describes the positions of poles of the physical 
amplitude $T_{1\bar{3}\to\bar{2}4}(s',t')$. Such poles are expected whenever 
an on-shell particle with appropriate mass $m^2=s'$ and angular momentum $l$ 
can be created in the collision of 1 and $\bar{3}$. Indeed, most Regge 
trajectories pass through known physical states with mass $m^2=t$ and 
angular momentum $\alpha(t)$. 

The Froissart bound implies that $\alpha(0)\le 1$ for all Regge 
trajectories. However, it was observed early on that a very good fit to
$pp$ and $p\bar{p}$ cross sections could be obtained assuming the dominance 
of a single pole with $\alpha(0)>1$. The corresponding trajectory is 
known as the pomeron trajectory. Donnachie and Landshoff found that 
a large set of different hadronic cross sections can be fitted with an 
intercept $\alpha(0)=1.08$~\cite{dl}. In spite of the power-like growth of 
Eq.~(\ref{rf}), the predicted cross sections are so small that the 
Froissart bound is not violated below the Planck scale. It is then argued 
that unitarity is not a serious problem at all realistic energies. However, 
one should keep in mind that analyses based on a pomeron trajectory with 
$\alpha(0)>1$ are, strictly speaking, not self-consistent in the framework 
of Regge theory. Therefore, it is likely that the rightmost singularity in 
the complex $l$ plane is not a single pole but a cut, in which case many 
of the results obtained in this framework are called into doubt. 

In connection with the observed strong rise of $\gamma^*p$ cross sections 
at HERA, there has recently been a lot of discussion of the possible need 
for a `second pomeron pole'~\cite{dl1} (see~\cite{lan} for an update). 
However, as will be discussed below, this effect can also be understood in 
QCD perturbation theory.

\subsection{Ideas in euclidean field theory}
As already mentioned, the best developed method for addressing 
non-perturbative problems in non-abelian gauge theories is the lattice. The 
numerical simulation of the path integral forces one to work in euclidean 
field theory and to derive Minkowski-space observables by analytic 
continuation. There are well-known methods for the hadron spectrum, certain 
decay processes, and for operators relevant to deep inelastic scattering
(DIS), but without the possibility to take the limit $\xbj\to 0$. However, 
no established procedure exists for the large-$s$ limit of hadronic cross 
sections. The fundamental difficulty becomes apparent by observing that the 
interesting dynamics resides in the soft fluctuations of the hadronic wave 
function, which, at $s\to\infty$ are localized on the light cone. 

An interesting idea to overcome this problem was put forward some time 
ago by Meggiolaro~\cite{meg} (see~\cite{meg1} for recent results). The 
approach rests on the well-known relation between the high-energy scattering 
of hadrons and correlators of light-like Wilson lines (which correspond to 
the trajectories of the constituent quarks)~\cite{nac}. Meggiolaro showed 
that the expectation value of two Wilson lines forming a certain hyperbolic 
angle in Minkowski space and the expectation value of two Wilson lines 
forming a certain angle in euclidean space are connected by analytic 
continuation in the angular variables. One may now hope to describe 
the limit of light-like minkowskian Wilson lines (i.e., the physical limit 
$s\to\infty$) on the basis of a lattice calculation of Wilson line 
correlators in the euclidean theory. Note also that a conceptually related 
but different euclidean approach to the small-$x$ limit of DIS was 
suggested in~\cite{hmn}. 

Of course, in the above proposal it is still not clear how to technically 
obtain the dependence on an angular variable with a precision that is high
enough for analytic continuation. The fundamental difficulty of this is 
obvious since the lattice breaks rotation invariance. This difficulty may
be deeply related to the second fundamental problem of lattice approaches 
to high-energy observables, namely, the vast discrepancy of the two 
scales $\Lambda_{QCD}$ and $s$. Such two-scale problems are difficult to
approach since, on the lattice, one is confined to the region of scales 
between lattice spacing and the size of the simulated box. 

A further new approach to the high-energy limit of QCD was proposed by Janik 
and Peschanski~\cite{jp} (see also~\cite{tan}). The authors suggest using the
AdS/CFT correspondence (also known as the Maldacena conjecture)~\cite{mal} 
to investigate high-energy scattering in non-abelian gauge theories. 
The AdS/CFT correspondence claims the equivalence of weakly coupled string 
theory in an Anti-de-Sitter (AdS) geometry with strongly coupled 
${\cal N}=4$ super Yang-Mills theory, which is a conformal field theory
(CFT), in 4-dimensional Minkowski space. Further, to make the connection 
to the realistic case of confining gauge theories, the authors use 
Witten's proposal~\cite{ew} that a confining gauge theory is dual to 
string theory in an AdS black hole background. As discussed above, the 
high-energy scattering of two dipoles can be calculated from the 
correlation function of two Wilson loops in the euclidean theory. Using 
AdS/CFT correspondence, the calculation of the latter can be reduced to a 
minimal surface problem in an AdS black hole background.

\subsection{The BFKL approach}
In brief, one could say that the BFKL (Balitsky-Fadin-Kuraev-Lipatov) method 
attempts to approach the rise of hadronic cross sections from the perturbative
side, by summing $\ln s$ enhancements appearing in higher orders of 
perturbation theory. Even though the technical realization is rather 
involved~\cite{bfkl} (see~\cite{fr} for a modern introductory text, 
~\cite{nlo} for results at next-to-leading order, and~\cite{rev} for 
recent reviews), the main physical idea is simple. Consider the scattering 
of two small (perturbative) dipoles (Fig.~\ref{bfkl}). At leading order, the 
total cross section is determined by one-gluon exchange. At next-to-leading 
order, a gluon can be radiated into the final state. The phase space open to 
this gluon is limited by the rapidities of the two colliding dipoles and 
grows with $s$. This leads to a $\ln s$ enhancement of the total cross
section. More final state gluons give rise to higher powers of $\ln s$. 
Keeping only the dominant terms, the whole series can be summed, giving rise 
to a cross section $\sigma\sim\alpha_s^2\,s^{\alpha_{BFKL}-1}$, where the 
`BFKL intercept' is given by $\alpha_{\sss BFKL}=1+12(\ln 2)\alpha_s/\pi$. 

\EPSFIGURE[ht]{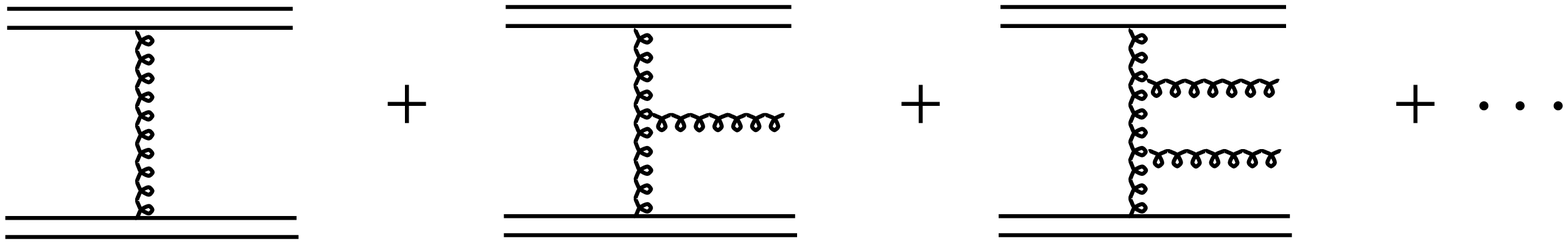,width=11cm}{Two dipoles scattering via one-gluon 
exchange. Gluons can be radiated into the final state.\label{bfkl}}

The $\ln s$ enhancement of higher orders in perturbation theory, which
underlies this result, is of fundamental importance because it represents 
our only lagrangian-based derivation of a growing high-energy cross section. 
Nevertheless, the BFKL method in its present form is far from answering the
fundamental question about the asymptotic high-energy behaviour. Firstly, 
it is not applicable to realistic hadrons because there is no hard scale 
justifying the perturbative method. Further, even if one restricts oneself 
to fictitious small-dipole (`onium-onium' -- cf.~\cite{mue}) collisions,
BFKL does not provide the answer at $s\to\infty$. On the one hand, it 
clearly violates the Froissart bound. On the other hand, it violates 
unitarity at fixed impact parameter. This can be understood by interpreting 
the BFKL gluon radiation as a growing `blackness' of the target, which is 
then probed by the projectile dipole. However, such an unlimited growth 
is impossible since the probability for the projectile to interact with the 
target at a given impact parameter eventually reaches unity. 

On the technical side, the two above difficulties are related to the 
problems of higher-order corrections and of infrared diffusion (see, 
e.g.~\cite{blv}). To understand the first problem, observe that even though 
BFKL sums terms $\sim\alpha_s^2(\alpha_s\ln s)^n$ for all $n$, one has all 
reason to expect the sum of contributions $\sim\alpha_s^4(\alpha_s\ln s)^n$ 
(e.g., `double pomeron exchange') to grow faster as $s\to\infty$. More 
generally, one expects that even the extension of BFKL to any finite higher 
order will not provide the true high-energy asymptotics of the cross section.
To understand the second problem, observe that the gluon ladder (the 
beginning of which is shown in Fig.~\ref{bfkl}) `knows' about the hard scale 
of the two scattering dipoles only via its ends. As $s$ increases, longer 
and longer gluon ladders become important and one sees numerically that the 
IR region starts to dominate the inner-rung momentum integrals.

Much of the recent work in BFKL physics has focussed on the proper 
understanding and implementation of the NLO corrections~\cite{nlo} which, 
naively, appeared to be extremely large, and on tests of BFKL dynamics at 
the available colliders. According to~\cite{blm}, the use of BLM scale 
setting improves the NLO situation dramatically. In a different approach, 
the improved small-$x$ evolution of~\cite{ccs}, which incorporates the 
next-to-leading order BFKL kernel as well as renormalization group 
constraints on the relevant collinear limits, stabilizes the resummed 
results. For recent discussions of IR diffusion and of the relevancy of BFKL 
to the high-energy limit of DIS the reader is referred to refs.~\cite{ctm} 
and~\cite{ccsx,abf}. Other theoretical directions include the further 
study of reggeization (see, e.g.~\cite{ewe}) and the development 
of a reggeon field theory approach~\cite{lip}. Being, at least in principle, 
very close to the ideal small-dipole case, the $\gamma^*\gamma^*$ cross 
section is an interesting testing field for BFKL methods. However, recent 
experimental data~\cite{ggvir} and theoretical analyses~\cite{cdft} show 
that it is difficult to identify any direct evidence for BFKL dynamics in 
$\gamma^*\gamma^*$ collisions at presently available energies. Note 
furthermore that recently total cross section measurements for the collision 
of two real photons have been extended to very high energies and that a 
faster rise than expected on the basis of soft pomeron parameterizations has 
been observed~\cite{ggrea}. 

Before closing this very brief section on BFKL, I would like to reiterate
that most recent developments in this field are to be considered as work
in perturbative QCD and that direct contact to the fundamental problem of 
the high-energy asymptotics is difficult to make. Therefore, in spite of 
its great interest in its own rights, the physics of BFKL lies somewhat 
outside the main line of development of this review.

\subsection{The method of high gluon densities}\label{hgd}

\EPSFIGURE[ht]{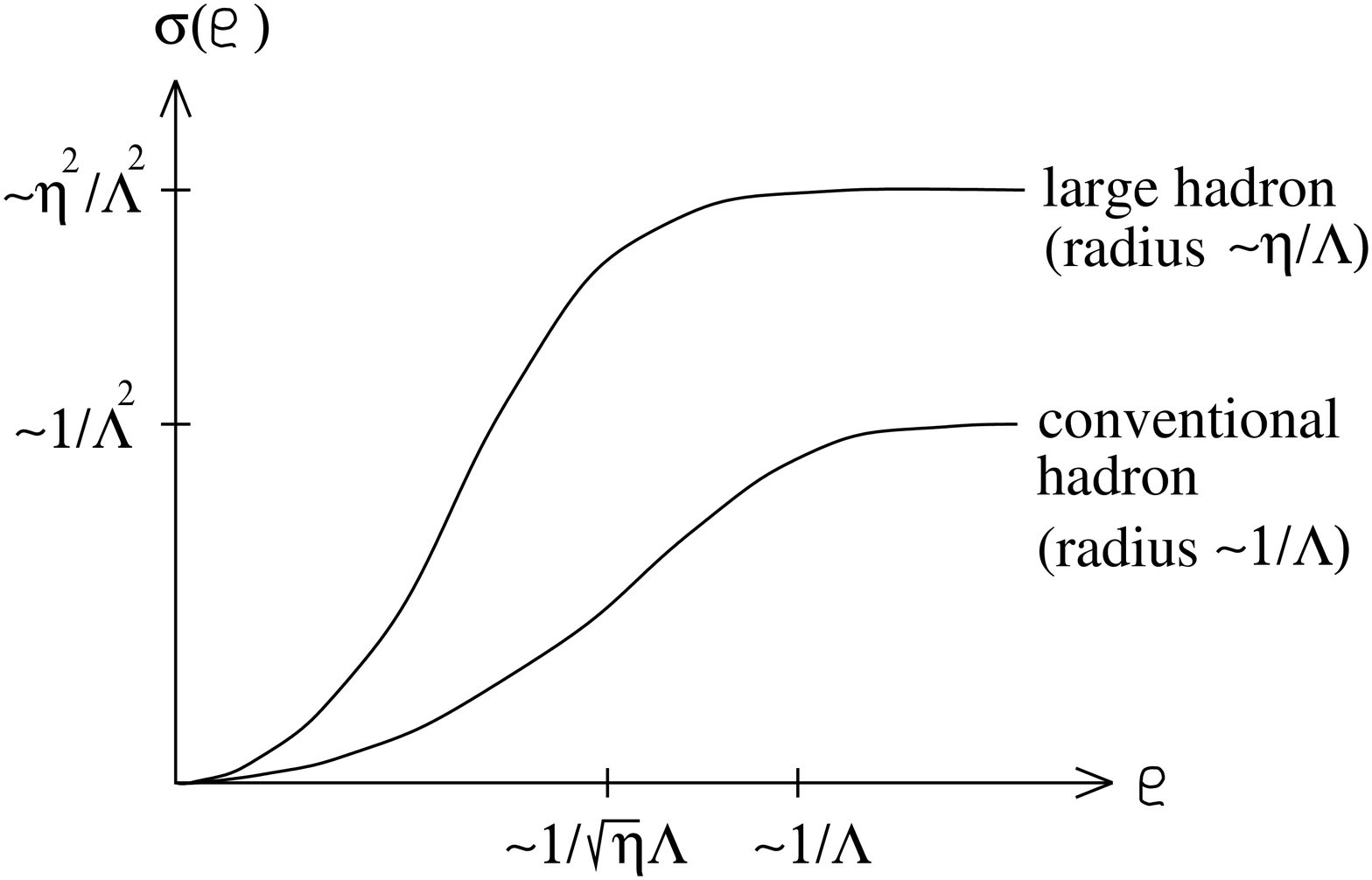,width=8.5cm}{Qualitative behaviour of the dipole
cross section $\sigma(\varrho)$ (here $\Lambda\sim\Lambda_{QCD}$).
\label{sigma}}

The gluon ladder of BFKL (see Fig.~\ref{bfkl}) can be interpreted as the 
consecutive radiation of gluons by the target hadron, so that the growth 
of the cross section corresponds to a growth of the target gluon 
distribution with increasing $1/x$. This small-$x$ enhancement of 
gluon distributions is also present in the more familiar DGLAP evolution of
parton distributions~\cite{dglap}. As already mentioned before, the
growth is limited by unitarity at any given point in impact space. The 
physical mechanism taming the growth is believed to be the recombination 
of partons~\cite{glr}. Thus, the region of high gluon densities appears 
to be highly relevant to the question of how hadronic cross sections behave 
at very high energies. 

A lot of interest has been drawn to the the region of high gluon densities 
since the McLerran-Venugopalan approach has offered the perspective of a 
new hardness scale and thus perturbative calculability in this region. The 
original proposal~\cite{mv} dealt with large nuclei, where the new hardness 
scale can be understood by boosting the nucleus to very large energy and 
observing that the thickness of the target translates to a high density 
(i.e., small transverse separation) of gluons in impact parameter space. 
This hardness scale can also be derived by considering small-$x$ DIS off a 
large nucleus in its rest frame~\cite{hw} (cf. also~\cite{jkmw,km}). In this 
approach, the total cross section is given by the convolution of the 
$\gamma^*$ wave function (characterizing the probability of a virtual 
photon to fluctuate into a $q\bar{q}$ pair of transverse size $\varrho$) 
and the dipole cross section $\sigma(\varrho)$ (characterizing the probability
of this dipole to interact with the target). For conventional hadrons,
even at high photon virtualities $Q^2$, dipole sizes up to $\sim 1/\Lambda$
contribute (the region of the `knee' in Fig.~\ref{sigma}). For very large 
hadrons, one can see by purely geometrical arguments that the region of 
the knee is shifted to smaller values of $\varrho$, the relevant scale 
being $\sqrt{\eta}\Lambda$. Thus, the non-perturbative region is 
parametrically suppressed in the total cross section. Intuitively, this 
effect can be explained by saying that the largest dipoles that 
contribute are those for which saturation sets in, and that in large (thick) 
targets saturation sets in already for small dipoles.

To make the connection to the high-energy (small-$\xbj$) limit, one now 
assumes that in the above arguments the thickness of the large target can 
be replaced by the extreme opacity that a usual hadronic target develops
when probed by a very energetic projectile. In other words, since the knee 
in Fig.~\ref{sigma} moves to the left in the small-$\xbj$ limit, saturation 
becomes a phenomenon that is, at least in principle, accessible in 
perturbation theory. Indeed, starting with~\cite{jkmw,jklw} (see~\cite{bal}
for earlier closely related work), over the last years impressive progress 
has been made in deriving and analysing the evolution of the gluon 
distribution within the McLerran-Venugopalan or high-density approach. In 
this framework, it is convenient to think of the hadron colour field as being
characterized by the expectation value of Wilson lines penetrating the 
hadron. In the simplest case, these are the two Wilson lines corresponding to
the $q\bar{q}$ pair into which the $\gamma^*$ fluctuates (cf. the quantity 
$\sigma(\varrho)$ above). To understand the energy dependence, one needs to 
consider an arbitrary number of such Wilson lines and to calculate the 
evolution of the generating functional for the corresponding expectation 
value in the target colour field. (Note that this can be understood as a
renormalization group equation in which more and more energetic gluon 
field components are integrated out.) Even assuming the validity of 
perturbation theory throughout, the problem at hand is formidable. 

\EPSFIGURE[ht]{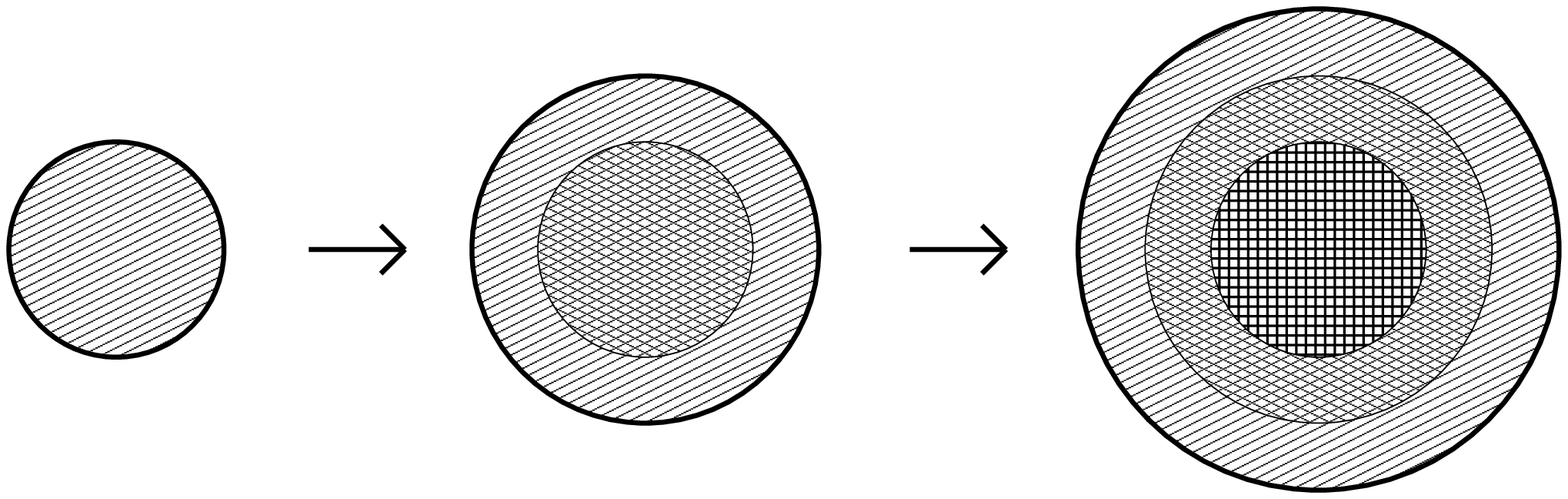,width=8.5cm}{Qualitative picture of the high-energy 
evolution of a hadronic target in impact parameter space.\label{disk}}

Recent work in the field is focussing on the full understanding of the 
above complicated non-linear evolution equation and on first attempts at 
physical applications~\cite{kmw,kov,wei,cgc,bb,kw}. In particular, the 
reader is referred to~\cite{kmw} for a comparative analysis of the methods 
employed in Refs.~\cite{jklw},~\cite{bal}, and~\cite{kov}. Progress in 
solving the evolution equation using functional Focker-Planck methods has 
been reported in~\cite{wei}. The application of the high gluon density 
regime, called `colour glass condensate' in~\cite{cgc}, to small-$x$ DIS has 
been discussed, e.g., in~\cite{bb,kw}. 

However, one now has to ask how much closer the above developments bring us 
to a solution of the fundamental problem -- the high-energy limit of hadronic 
cross sections. To discuss this question, consider the change of the impact 
parameter space picture of a hadron with increasing energy,~Fig.~\ref{disk}. 
Perturbation theory tells us that, at any given impact parameter, the gluon 
density (or, equivalently, the interaction probability of an energetic small 
colour dipole) increases with energy. In Fig.~\ref{disk}, this is symbolized 
by increasing blackness. For given impact parameter and given dipole size, 
this growth is limited by unitarity. Technically, this is implemented by 
including non-linearities into the evolution equation -- this is precisely 
the program of the McLerran-Venugopalan approach sketched above. Thus, one 
may indeed hope that this program will lead to a quantitative understanding 
of how, in the high energy limit, the target becomes completely black at 
any impact parameter. However, if total hadronic cross sections continue 
to grow asymptotically as $s\to\infty$, then this growth has to come from 
an effective transverse expansion in impact parameter space. As illustrated 
in Fig.~\ref{disk}, this expansion is governed by the dynamics near the edge
of the disk, a region where gluon densities are not high and the 
applicability of the McLerran-Venugopalan method is not justified in any
obvious way. In my opinion, it is this `transverse' dynamics, i.e., the 
expansion of the target disk into the previously `white' region, which 
represents the main challenge to the method of high gluon densities.

\section{Small-$x$ diffraction}
\subsection{Diffractive DIS as a tool to study high-energy QCD}

The small-$x$ limit of DIS (deep inelastic scattering) became experimentally 
viable only with the advent of the electron-(or positron-)proton collider 
HERA. The large centre-of-mass energy of the $ep$ collision, $\sqrt{s}\simeq 
300$ GeV, allows for a very high hadronic energy $W$ (the cms energy of the 
$\gamma^*p$ collision) and thus for the observation of events with both very 
high photon virtuality $Q^2$ and very small $\xbj=Q^2/(Q^2+W^2)$. 

\EPSFIGURE[ht]{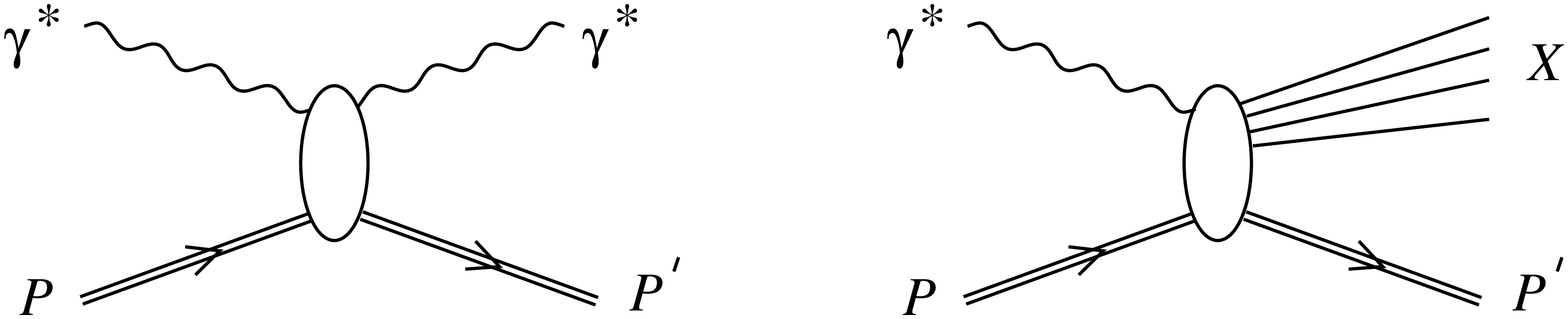,width=10cm}{Forward Compton scattering and 
diffractive electroproduction.\label{dep}}

Loosely speaking, diffraction is the subset of DIS characterized by a 
quasi-elastic interaction between virtual photon and proton. A particularly 
simple definition of diffraction is obtained by demanding that, in the 
$\gamma^*p$ collision, the proton is scattered elastically. Thus, in 
diffractive events, the final state contains the scattered proton with 
momentum $P'$ and a diffractive hadronic state $X_M$ with mass $M$ (see 
the r.h. side of Fig.~\ref{dep}). 

Figure~\ref{dep} illustrates that diffractive DIS can be understood as a 
very special type of hadronic high-energy process. Inclusive small-$x$ 
DIS, i.e., the total high-energy $\gamma^*p$ cross section, is linked by the 
optical theorem to forward Compton scattering (l.h. side of Fig.~\ref{dep}). 
This is already a very interesting process since one has the additional 
variable $Q^2$ at one's disposal. In this sense, diffraction (r.h. side of 
Fig.~\ref{dep}) is an even more interesting hadronic high-energy reaction 
since one now can also use the momentum transfer $t$ and the details of the 
hadronic final state $X$ to control the interaction between the two 
colliding objects. Given our ability to measure small-$x$ diffractive cross 
sections with high precision at the HERA accelerator, diffractive DIS can
be used as a very powerful and versatile tool for the study of the 
high-energy behaviour of QCD, which significantly extends the physics 
capacity of hadron-hadron colliders. For recent reviews of diffractive DIS, 
the reader is referred to~\cite{revd}.

\subsection{The aligned jet model and its modern versions}\label{ajmmod}
The aligned jet model~\cite{bk} is based on a qualitative picture of DIS 
in the target rest frame, where the incoming virtual photon can be 
described as a superposition of partonic states. The large virtuality 
$Q^2$ sets the scale, so that states with low-$p_\perp$ partons, i.e., 
aligned configurations, are suppressed in the photon wave function. 
However, in contrast to high-$p_\perp$ configurations, these aligned 
states have a large interaction cross section with the proton. Therefore, 
their contribution to DIS is expected to be of leading twist (of leading 
order in $1/Q^2$). Since the above low-$p_\perp$ configurations represent 
transversely extended, hadron-like objects, which have a large elastic 
cross section with the proton, part of this leading twist contribution is 
diffractive. This very simple picture explains on a qualitative level the 
large cross section for diffractive or `rapidity gap' events the observation 
of which caused significant excitement in the beginning of the HERA 
era~\cite{rge}. 

The above intuitive picture was implemented in the framework of perturbative 
QCD in~\cite{wf}, where the perturbative fluctuation of the $\gamma^*$ into 
a $q\bar{q}$ pair was considered as the dominant process and the subsequent 
colour singlet exchange between the proton and the $q\bar{q}$ pair 
was realized by two gluons. A further essential step is the inclusion of 
higher Fock states in the photon wave function. In the framework of two 
gluon exchange, corresponding calculations for the $q\bar{q}g$ state were 
performed in~\cite{nz}. The main shortcoming of the two-gluon approach is 
the problem of justifying perturbation theory. As should be clear from 
the qualitative discussion of the aligned jet model, the diffractive 
kinematics is such that the $t$ channel colour singlet exchange does not 
feel the hard scale of the initial photon. Thus, more than two gluons can 
be exchanged without suppression by powers of $\alpha_s$. 

This problem was systematically addressed in the semiclassical 
approach~\cite{bh,bmh}, where the interaction with the target was modelled 
as the scattering off a superposition of soft colour fields. 
In the high-energy limit, the eikonal approximation can be used
to calculate the scattering of the energetic partons of the $\gamma^*$ 
fluctuation. Diffraction occurs if both the target and the partonic 
fluctuation of the photon remain in a colour singlet state. Thus, both the 
diffractive and inclusive DIS cross section can be calculated if a model 
for the wave functional of the proton is provided. A simple 
phenomenologically successful model for the proton colour field, which 
is based on the Glauber formula justified by the large-target approximation, 
was developed in~\cite{bgh}. 

Both the two-gluon-exchange and the semiclassical approach contain, as
a first step, the perturbative calculation of the partonic $\gamma^*$ wave 
function, i.e., of the transition amplitude from the $\gamma^*$ to the 
$q\bar{q}$ or $q\bar{q}g$ state. In the simpler $q\bar{q}$ case, this 
wave function is then convoluted with the amplitude for the elastic 
interaction of the colour dipole and the target hadron. At $t=0$, this 
amplitude is determined by the dipole cross section $\sigma(
\varrho)$ (cf.~Sect.~\ref{hgd}). Thus, one can now discuss different 
models for the target on the basis of $\sigma(\varrho)$. In particular,
$\sigma(\varrho)$ can be modelled by perturbative two-gluon exchange with 
an ad-hoc IR cutoff or by the expectation value of two Wilson lines in the 
proton state (which is IR finite for a finite-size proton). 

As far as $\sigma(\varrho)$ is concerned, certain relatively 
model-independent general statements can be made. The firmest one is 
probably the well-known relation to the inclusive gluon 
distribution $xg(x,\mu^2)$~\cite{fms}, 
\beq
\sigma(\rho)=\frac{\pi^2}{3}\alpha_s[xg(x,1/\rho^2)]\rho^2+{\cal O}
(\rho^4)\,,
\eeq
valid at small $\rho$. Note that this implies a dependence of $\sigma(\rho)$
on the $x$ (or, equivalently, on the energy of the projectile $q\bar{q}$ 
pair) since $xg(x,\mu^2)$ is known to increase with $1/x$ in the small-$x$
limit. 

An intuitively obvious but quantitatively less 
clear statement is that this quadratic rise of $\sigma(\rho)$ will 
eventually be tamed by non-perturbative effects, .i.e., a saturation of 
$\sigma(\rho)$ has to occur at large $\rho$. In the case of a very large 
target, this saturation sets in when $\rho$ is still in the perturbative 
domain and a Glauber-type formula for $\sigma(\rho)$ can be 
derived~\cite{bgh}. For a realistic hadron, the situation is less clear and 
the functional form of the saturation is unknown. Furthermore, one has to 
worry that saturation occurs at non-perturbative values of $\rho$, where the 
dipole picture itself is questionable (see below). 

A very simple parameterisation of $\sigma(\rho)$, which takes into account 
the above generic features and which has been very popular recently, was
suggested in~\cite{gw}. It reads 
\beq
\sigma(\rho,x)=\sigma_0 \left\{1-\exp\left(-\frac{\rho^2}{4R_0^2(x)}\right)
\right\}\,,
\eeq
with $R_0=(1/$GeV$)(x/x_0)^{\lambda/2}$, i.e., it assumes a power-like 
growth of $xg(x,\mu^2)$ and a Glauber-type shape of saturation. The strength 
of this parameterisation lies in its simplicity and in the fact that it 
captures several important qualitative features of the dipole cross section. 
One shortcoming is that, in this parameterisation, the level at which 
$\sigma(\rho)$ saturates is given by $\sigma_0$ and is therefore independent 
of the energy. This contradicts our experimental knowledge that 
soft hadronic cross sections grow with $s$. An interesting result of the 
application of the above parameterisation to diffractive and inclusive DIS 
is that both processes have a similar energy dependence. This remarkable 
scaling behaviour was previously pointed out in~\cite{buch} and derived 
from very general considerations in the semiclassical framework 
in~\cite{bh}. 

Unfortunately, there is one rather fundamental criticism that applies to
all the above calculations based on the perturbative transition of the 
$\gamma^*$ to a set of partons and their subsequent interaction with the 
target proton. The problem is that the bulk of diffraction comes from 
partonic configurations, in particular $q\bar{q}$ pairs, which are not
parametrically small on a scale of $\Lambda_{QCD}$. Therefore, strictly 
speaking there is no reason to neglect non-perturbative gluonic 
interactions between quark and antiquark, and even worse, the interaction 
of the non-perturbative gluon field between the quarks with the gluon 
field of the target. To the best of my knowledge, the only case where this 
problem is under control is diffractive DIS of an optically very thick 
target described above. The required optical thickness can also arise in
diffraction of protons at extremely high energies, but the applicability of 
this argument in the HERA domain is less than obvious.

\subsection{Diffractive parton distributions}
The conventional partonic interpretation of inclusive DIS appears to be 
most natural in the Breit frame (the frame where the photon momentum has 
the form $q=(0,0_\perp,Q)$ and the proton energy becomes very large as 
$x\to 0$). Viewing diffractive DIS in analogy to inclusive DIS, one arrives 
at a picture that is very different from the target rest frame picture of 
Sect.~\ref{ajmmod}. In this approach, the concept of 
fracture functions~\cite{vt} or, more specifically, the diffractive parton 
distributions of~\cite{bs} provide a framework firmly rooted in perturbative 
QCD. Recall that conventional parton distributions can be considered as 
probabilities for finding a parton with a certain momentum fraction in the 
fast moving proton. In short, diffractive parton distributions are 
conditional probabilities. A diffractive parton distribution 
$df^D_i(y,\xi,t)/d\xi\,dt$ describes the probability of finding, in a fast 
moving proton, a parton $i$ with momentum fraction $y$, under the additional 
requirement that the proton remains intact while being scattered with 
invariant momentum transfer $t$ and losing a small fraction $\xi=x_{\pom}$ 
of its longitudinal momentum. Thus, the corresponding $\gamma^*p$ cross 
section can be written as~\cite{bs1}
\beq
\frac{d\sigma(x,Q^2,\xi)^{\gamma^*p\to p'X}}{d\xi}=\sum_i\int_x^\xi dy\,
\hat{\sigma}(x,Q^2,y)^{\gamma^*i}\left(\frac{df^D_i(y,\xi)}{d\xi}\right)\, ,
\label{sx}
\eeq
where $\hat{\sigma}(x,Q^2,y)^{\gamma^*i}$ is the total cross section for 
the scattering of a virtual photon characterized by $x$ and $Q^2$ and a 
parton of type $i$ carrying a fraction $y$ of the proton momentum. The 
above factorization formula holds in the limit $Q^2\to\infty$ with $x$, 
$\xi$ and $t$ fixed. Factorization proofs were given in~\cite{gtv} in the 
framework of a simple scalar model and in~\cite{cfa} in full QCD. 

\EPSFIGURE[ht]{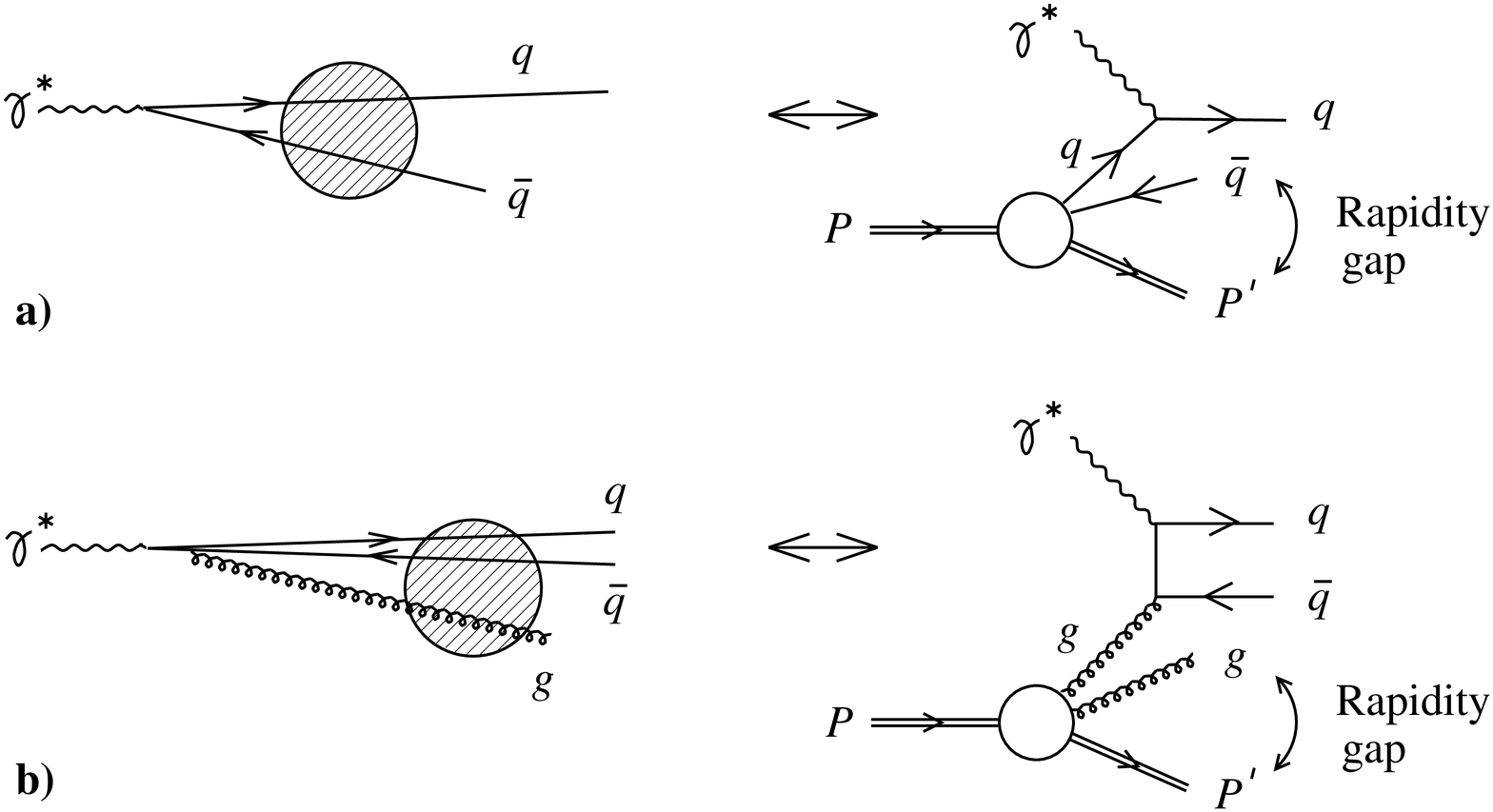,width=10cm}{
Diffractive DIS in the proton rest frame (left) 
and the Breit frame (right); asymmetric quark fluctuations correspond to 
diffractive quark scattering, asymmetric gluon fluctuations to diffractive 
boson-gluon fusion.\label{f2d}}

\EPSFIGURE[ht]{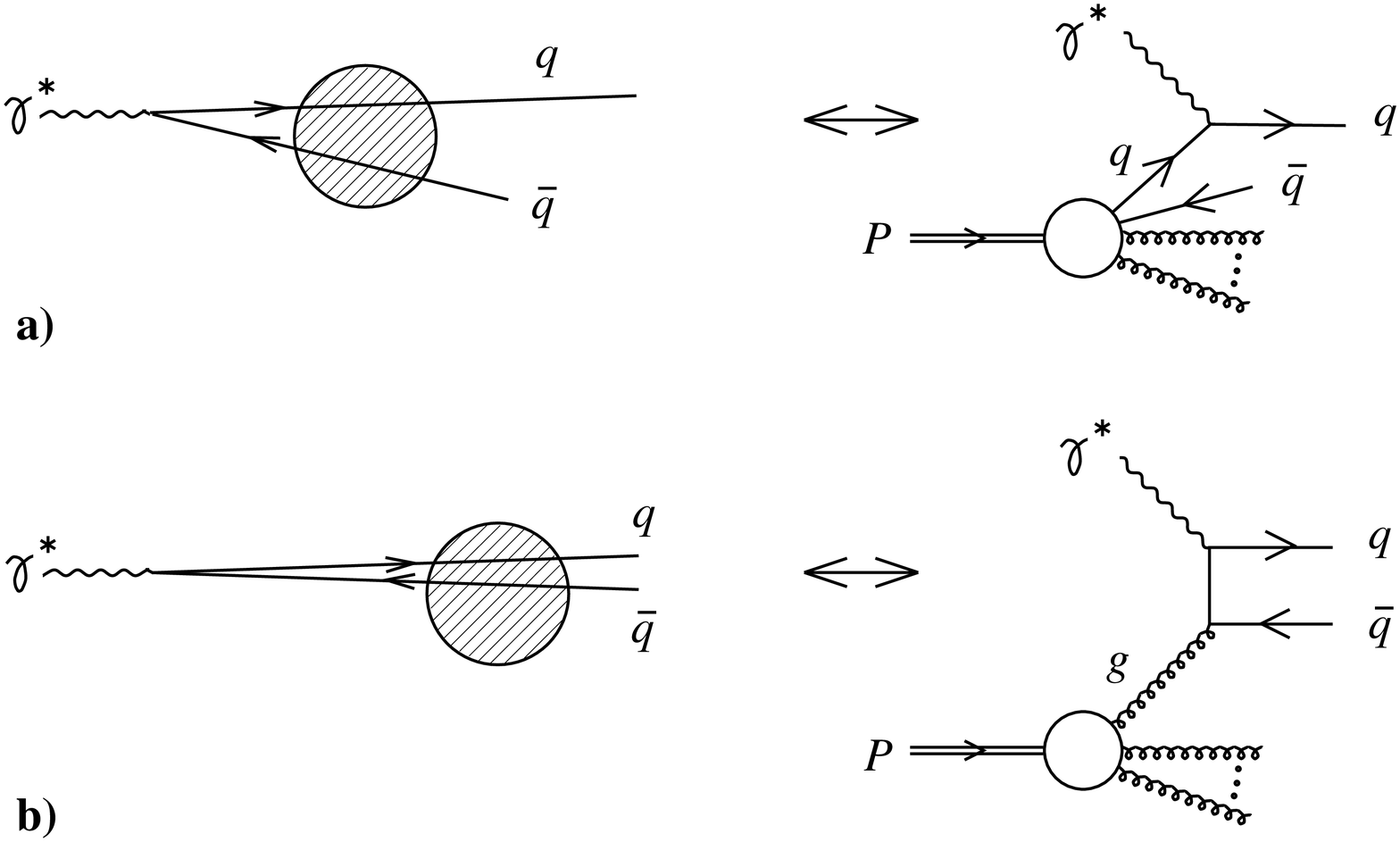,width=10cm}{
Inclusive DIS in the proton rest frame (left) 
and the Breit frame (right); asymmetric fluctuations correspond to quark 
scattering (a), symmetric fluctuations to boson-gluon fusion (b).\label{f2}}

As in inclusive DIS, there are infrared divergences in the partonic cross 
sections and ultraviolet divergences in the parton distributions. Thus, a 
dependence on the factorization scale $\mu$ appears both in the parton 
distributions and in the partonic cross sections. The claim that 
Eq.~(\ref{sx}) holds to all orders implies that these $\mu$ dependences 
cancel, as is well known in the case of conventional parton distributions. 
Therefore, the diffractive distributions obey the usual DGLAP evolution 
equations~\cite{bs1,blu}. 

\EPSFIGURE[ht]{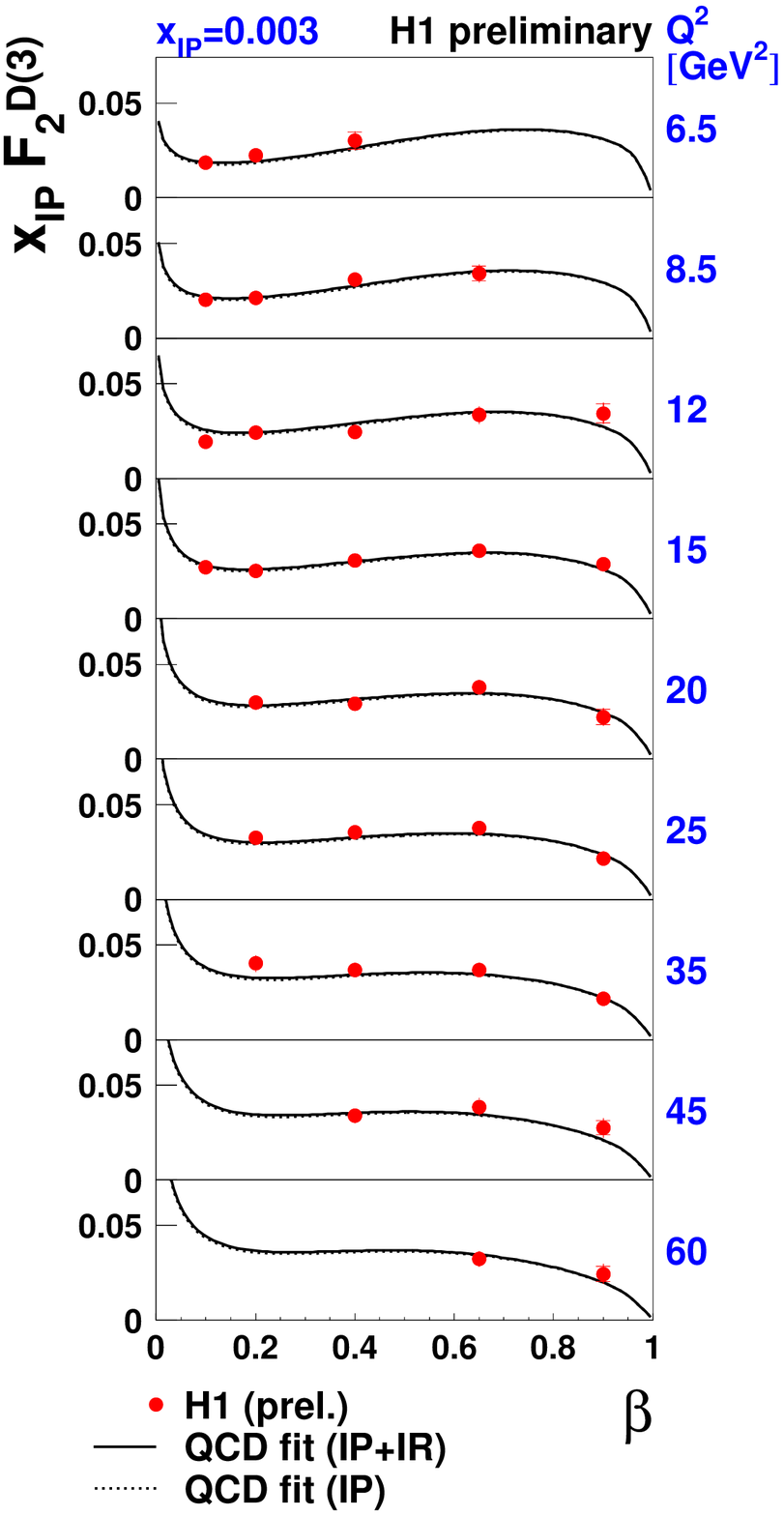,width=6cm}{The measured diffractive structure 
function, plotted as $x_{\pom}F_2^{D(3)}(\beta,Q^2,x_{\pom})$ ({\it red data 
points}), as a function of $\beta$ for various values of $Q^2$ and a fixed 
value $x_{\pom}=0.003$. Also shown is the result of a QCD fit to the data as
described in the text of conference paper 808. The solid curves correspond 
to the sum of pomeron and leading reggeon exchange contributions, whereas 
the dotted curves indicate the contribution from pomeron exchange only.
(Figure from H1 conference paper 808~\cite{new}).
\label{newm}}

The connection of the above picture of diffractive and inclusive DIS with 
the previously discussed aligned jet or target rest frame picture is 
illustrated in Figs.~\ref{f2d} and~\ref{f2}. Consider, for example, the 
diffractive quark distribution (Fig.~\ref{f2d}a). The leading twist 
contribution arises from fluctuations where one of the two quarks carries 
only a small longitudinal momentum fraction of the $\gamma^*$. Boosting 
to the Breit frame, one finds that this less energetic antiquark can be 
interpreted as a quark coming from the diffractive quark distribution of
the proton. Putting this and the other subprocesses in Figs.~\ref{f2d} 
and~\ref{f2} into equations, one finds that the diffractive and inclusive 
parton distributions can be expressed in terms of expectation values of 
Wilson loops in the proton colour field (or, similarly, in terms of the 
colour dipole cross sections $\sigma(\rho)$)~\cite{h,bgh}. One can then 
perform phenomenological analyses of inclusive and diffractive DIS on the 
basis of a model for the proton at some low scale $Q_0^2$ using conventional 
DGLAP evolution to make predictions at all $Q^2\geq 
Q_0^2$~\cite{bgh,hks,gw1}. 

Historically, the concept of diffractive parton distributions has a 
predecessor in the partonic interpretation of the pomeron~\cite{is}. In this
approach, the quasi-elastic high-energy scattering of photon fluctuation 
and proton is interpreted in terms of pomeron exchange and it is assumed 
that the pomeron can, like a real hadron, be characterized by a parton 
distribution. This distribution is assumed to factorize from the 
pomeron trajectory and the pomeron-proton-proton vertex, which are both 
obtained from the analysis of purely soft hadronic reactions. The problem 
with this approach is the lacking justification of the pomeron idea and 
the factorization assumption in QCD. Furthermore, the observed energy 
dependence of diffractive DIS disagrees with the universal soft pomeron
expectation, and the universality between diffraction in DIS and in 
hadron-hadron collisions, which is expected in this approach, is not 
observed~\cite{alv}. As reported at this conference~\cite{tim}, a unified
description of diffractive DIS and hadronic diffraction at the Tevatron
can be achieved in Monte-Carlo models based on soft colour 
exchange~\cite{ier}. For other approaches to this difficult problem see,
e.g.,~\cite{surv}.

\subsection{Hard colour dipole exchange}
Diffractive processes where the $t$ channel colour singlet exchange is 
governed by a hard scale include the electroproduction of heavy vector 
mesons~\cite{rys}, electroproduction of light vector mesons in the case of 
longitudinal polarization~\cite{bro} or at large $t$~\cite{fry}, and virtual 
Compton scattering (the process $\gamma^*p\to\gamma p'$)~\cite{mea,ji,dmue}. 
In the leading logarithmic approximation, the relevant two-gluon form 
factor of the proton can be related to the inclusive gluon 
distribution~\cite{rys}. Accordingly, a very steep energy dependence of the 
cross section, which is now proportional to the square of the gluon 
distribution, is expected.

To go beyond leading logarithmic accuracy, the non-zero momentum transferred 
to the proton has to be taken into account. This requires the use of 
skewed parton distributions (see~\cite{mea} and refs.~therein), which 
were discussed in~\cite{ji} within the present 
context. Although their scale dependence is predicted by well-known 
evolution equations, only limited information about the relevant input 
distributions is available (see, however,~\cite{fnf} for possibilities of 
predicting the non-forward from the forward distribution functions).
For the application of an NLO analysis to virtual Compton scattering
the reader is referred to~\cite{mf} and refs.~therein.

The perturbative calculations of meson electroproduction discussed above 
were put on a firmer theoretical basis by the factorization proof of 
\cite{cfs1}. Note, however, that the naive power counting expectation 
of the $Q^2$ dependence of cross sections is significantly modified by
the anomalous dimension of the gluon distribution (cf.~\cite{mrt}, where
even the notoriously difficult ratio of longitudinal and transverse cross
section is explained in a perturbative calculation). An interesting new 
field, closely related to exclusive vector meson production, are the 
semi-exclusive processes discussed in~\cite{sep}. 

In connection with inclusive diffraction, it is interesting to point out
that $F_2^{D(3)}(\beta,Q^2,\xi)$ at $\beta\to 1$ is dominated by hard colour 
singlet exchange and can therefore be calculated from the skewed gluon 
distribution~\cite{ht} (see also~\cite{bekw}). Given the growing precision 
of the data, this perturbative character of $F_2^D$ at large $\beta$ will
be essential for a full quantitative understanding of inclusive diffraction.

\subsection{New precision data from HERA}
\EPSFIGURE[ht]{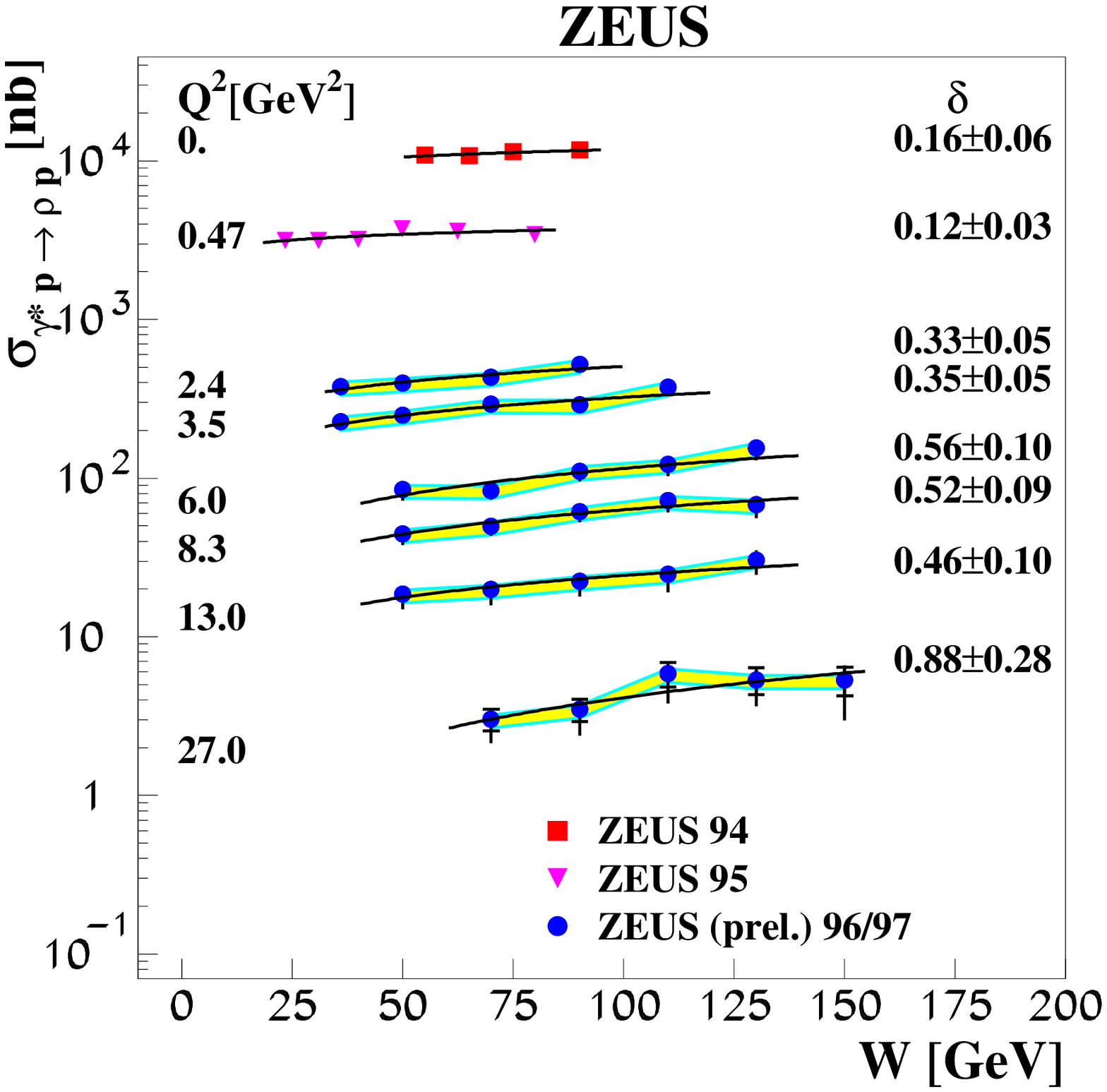,width=10cm}{$W$ dependence of the cross section 
$\sigma(\gamma^*p\to\rho^0 p)$ for various $Q^2$ values as denoted in the 
figure. The data for $Q^2<1$GeV$^2$ obtained in previous ZEUS 
measurements~\cite{pzeus} are shown for completeness. The dashed lines 
represent the results of fitting $\sigma\sim W^\delta$ at each $Q^2$ value 
and the results of the fit are shown in the figure. The shaded area indicates 
additional normalization uncertainties due to proton dissociation background
(Figure from ZEUS conference paper 594~\cite{mey}).
\label{mey}}

The purpose of this section is to emphasize that new small-$x$ data from 
HERA is at present transforming diffractive DIS into a precision field. 
This transformation will be taken even further once the data from the
HERA high-luminosity run starts to become available. As an illustration of 
the new quality of the data, consider the diffractive structure function 
measurement shown in Fig.~\ref{newm} (for previous results see~\cite{f2d}). 
To understand this quantity, recall that, in addition to the conventional 
kinematic variables of DIS, $Q^2$ and 
$\xbj$, the diffractive process is characterized by $M$, the mass of the 
diffractive final state $X$. Alternatively, the variables 
$\beta=Q^2/(Q^2+M^2)$ or $\xi=x_{I\!\!P}=x/\beta$ can be used. Now, 
$F_2^{D(3)}(x,Q^2,\xi)$ is defined precisely as $F_2(x,Q^2)$, but on the 
basis of a cross section that is differential in $\xi$ as well as in $x$ and 
$Q^2$. Alternatively, one can replace $x$ by $\beta$ and write $F_2^{D(3)}
(\beta,Q^2,\xi)$. The $Q^2$ evolution of the $\beta$ dependence, an 
essential ingredient of the method of diffractive parton distributions, is 
clearly visible in Fig.~\ref{newm}. Furthermore, the $\beta$ dependence is 
measured so well that one can now really hope to use diffraction as a tool to
study the colour field of the proton. (Recall that, using the semiclassical 
framework, the $\beta$ dependence of the diffractive structure function is 
linked to the functional form of the expectation value of a Wilson loop in 
the proton state.)

Furthermore, as can be seen in Fig.~\ref{mey}, the energy dependence of 
vector meson electroproduction can now be measured in detail for different 
values of the photon virtuality $Q^2$. Since $Q^2$ is linked to the typical 
size of the $q\bar{q}$ dipole probing the proton field, one has yet another
tool to study the energy dependence of the dipole cross section or, in other 
terms, the energy evolution of the effective proton colour field. 
Figure~\ref{mey} shows the expected steeper energy dependence for small 
dipoles, where perturbative calculations are relevant. Note that it is an
important long-term goal to experimentally verify that the fast growth of 
small-dipole cross sections softens above a certain energy (saturation in 
energy). 

An interesting new result reported at the conference~\cite{gol} concerns
the odderon exchange (a gluonic $t$ channel exchange with $C=P=-1$). Given 
the observed suppression of the odderon in $pp$ and $p\bar{p}$ cross 
sections, it was suggested in~\cite{ber} that the reason lies in the 
quark-diquark structure of the proton. If this is the case, then the 
diffractive production of pseudoscalar mesons at HERA with proton breakup 
in the final state should provide an ideal testing ground for the 
odderon. Employing a stochastic vacuum based approach in the description 
of the soft $t$-channel exchange~\cite{msv}, a prediction for, e.g., 
$\sigma_{\gamma p\to\pi^0X}$ was derived~\cite{ber}. However, the 
measurements show no trace of this and similar odderon induced 
reactions~\cite{gol} and thus the absence of the odderon remains mysterious. 

In the future, it will be very interesting to see precision measurements 
of the diffractive structure function based on a tagged leading proton. First
of all, such a measurement corresponds to the simplest and cleanest 
definition of diffractive DIS from the theoretical perspective. In 
particular, only in processes where the scattered proton is tagged can one 
be certain that the $\gamma^*$ partonic fluctuation was really 
probing the proton colour field and not some more complicated 
field corresponding to the transition between the proton and one of its 
excitations. Second, tagging the proton opens up the possibility of 
measuring the $t$ dependence of diffraction, i.e., $F_2^{D(4)}(\beta,Q^2,
\xi,t)$, and this has the potential of resolving the transverse form of
the target proton as seen at very high energy. This last possibility has 
already been exploited in~\cite{msm} in the context of elastic meson 
production. Note that new and interesting diffractive measurements with 
tagged leading proton have been reported at this conference~\cite{lps}.

\section{Conclusions and outlook}
The high-energy limit of hadronic cross sections is a longstanding and 
fundamental problem in our understanding of the known interactions. In 
my opinion, it is at present not clear from which direction a solution 
might eventually emerge. There exist only few conceptually clean, 
lagrangian-based ideas of how to approach this problem. Among those 
are attempts of a translation of the problem to euclidean field theory, with 
the aim to perform a lattice calculation, as well as attempts to make use of 
the non-perturbative understanding of non-abelian gauge theories obtained in 
the framework of the AdS/CFT correspondence. 

The only well-established QCD-based method of obtaining growing hadronic
cross sections is, at present, the summation of $\ln s$ enhanced diagrams 
in perturbation theory. However, even though very impressive technical 
progress in this framework has been made over the past few years
(NLO BFKL calculation, non-linear evolution equations at high gluon 
densities) it appears unlikely that the true high-energy asymptotics 
will become accessible with the methods available at present. 

On the phenomenological side, the HERA accelerator is providing a rich 
and fruitful testing ground for theoretical ideas and methods. In particular, 
the measurement of diffractive and inclusive DIS allows for the study of
hadronic reactions where one of the `hadrons', the partonic fluctuation of 
the $\gamma^*$, can be tuned to one's needs. The most recent HERA results 
on small-$x$ physics have reached an impressive precision, and future 
results from the HERA high-luminosity run will transform diffractive 
high-energy scattering into a precision field. One may hope that, in the 
more distant future, the realisation of TESLA$\times$HERA (THERA) will 
further expand the energy reach of the very promising, multifaceted
reaction channel $\gamma^*p$~\cite{thera}. 

We look forward to the forthcoming precision data, the careful analysis of 
which we hope will provide essential clues to finding a first-principles 
theoretical approach to the complicated problem of the high-energy limit of 
QCD cross sections.

\subsection*{Acknowledgements}
I would like to thank the organizers of EPS-HEP 2001 for a very enjoyable 
and interesting conference and for their invitation to present this review.

\end{document}